\documentclass{iau_FM}
\usepackage{natbib}
\usepackage{graphicx}

\newcommand{\Msun}{\ensuremath{\,{\rm M}_\odot}}                  
\newcommand{\Rsun}{\ensuremath{\,{\rm R}_\odot}}                  
\newcommand{\Teff}{\ensuremath{T_{\rm eff}}}                      
\newcommand{\kepler}{\textit{Kepler}}                             

\title[FM17.~~Pulsations in eclipsing binaries]{Double riches: \\ asteroseismology in eclipsing binaries}
\author[John Southworth]{John Southworth$^1$}
\affiliation{$^1$Astrophysics Group, Keele University, Staffordshire, ST5 5BG, UK\\email: {\tt astro.js@keele.ac.uk}}
\pubyear{2015}
\jname{Focus on Astronomy}
\editors{Simon Jeffery \& Joyce Guzik}

\begin{document} \maketitle 

\begin{abstract}
The study of eclipsing binaries is our primary source of measured properties of normal stars, achieved through analysis of light and radial velocity curves of eclipsing systems. The study of oscillations and pulsations is increasingly vital for determining the properties of single stars, and investigating the physical phenomena active in their interiors. Combining the two methods holds the promise of establishing stringent tests of stellar evolutionary theory, and of calibrating model-dependent asteroseismology with empirically measured stellar properties. I review recent advances and outline future work.
\end{abstract}

\firstsection


\section{Introduction}

The study of eclipsing binary star systems (hereafter EBs) has a long history. The eclipse hypothesis was proposed as an explanation of the periodic dimmings of the `demon star' Algol ($\beta$\,Persei) by John \citet{Goodricke1783}. The first empirical measurement of the masses and radii of two stars in an EB was that by \citet{Stebbins11apj} for $\beta$\,Aurigae; the numbers are close to modern values \citep{Me++07aa}. Shortly afterwards, \citet{Russell12apj} established a quantitative method for analysing light curves of eclipses.

\begin{figure}[t]
\begin{center}
\includegraphics[width=0.8\textwidth]{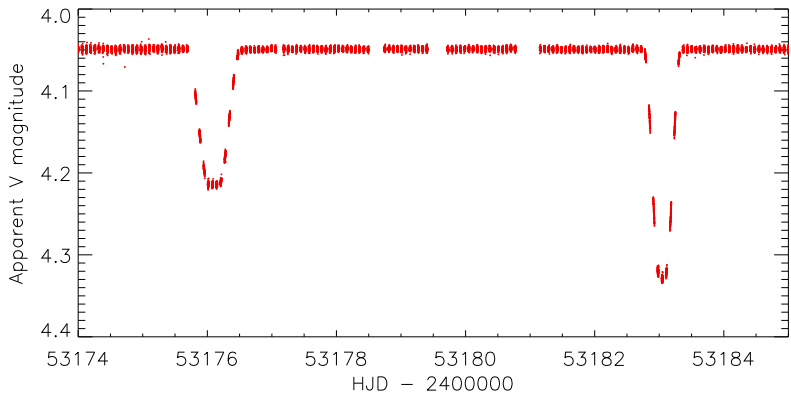}
\caption{WIRE satellite light curve of the EB $\psi$\,Centauri \citep{Bruntt+06aa}.}
\label{fig:pcen}
\end{center}
\end{figure}

\begin{figure}[t]
\begin{center}
\includegraphics[width=0.48\textwidth]{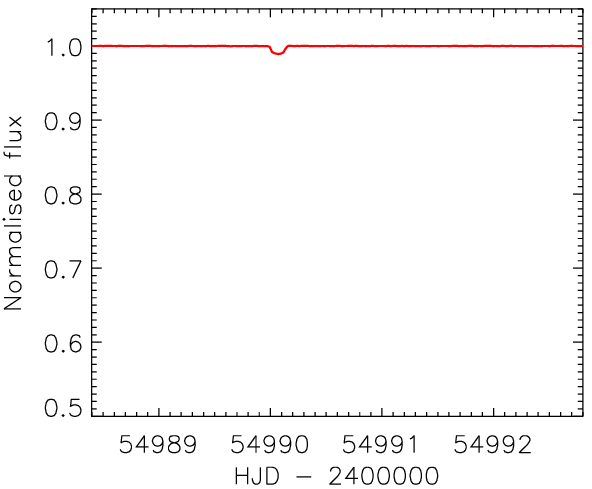}
\includegraphics[width=0.48\textwidth]{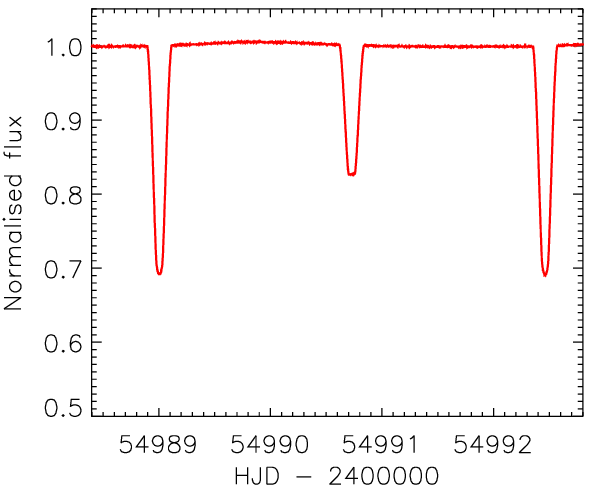}
\caption{\kepler\ satellite light curves of the transiting planetary system
Kepler-6 (left) and the EB KIC\,5288543 (right) on the same axis scales.}
\label{fig:comp}
\end{center}
\end{figure}

The era of high-quality space photometry began in 2006, with a light curve of $\psi$\,Centauri from the WIRE satellite \citep{Bruntt+06aa}. This star shows deep total eclipses on a 38.8\,d period (Fig.\,\ref{fig:pcen}), and prior to the serendipitous observations from WIRE was not known as either an eclipsing or spectroscopic binary despite its brightness ($V = 4.05$). The current era of vast photometric surveys has led to the discovery of thousands of new EBs, with extensive light curves being obtained particularly by surveys for transiting planets (e.g.\ TrES, HAT, WASP, CoRoT, \kepler, and in the future TESS and PLATO). Fig.\,\ref{fig:comp} compares the light curves of a transiting planetary system and EB: it is clear that any decent photometric survey for transiting planets is easily capable of detecting eclipsing binaries.

The importance of EBs lies in their amenability to detailed analysis. From fitting a light curve with a simple geometrical model one can determine the fractional radii of the two stars ($r_{\rm A} = \frac{R_{\rm A}}{a}$ and $r_{\rm B} = \frac{R_{\rm B}}{a}$ where $R_{\rm A,B}$ are the true radii of the stars and $i$ is the orbital inclination of the system). Spectroscopic radial velocity measurements for both stars allow the minimum masses ($M_{\rm A} \sin^3 i$ and $M_{\rm B} \sin^3 i$) and orbital separation ($a \sin i$) to be calculated. These quantities immediately yield measurements of the masses and radii of the two stars, which can be achieved empirically and to accuracies better than 1\% if the observational data are of good quality \citep[e.g.][]{Clausen+08aa}.

Obtaining the \Teff\ values of the stars -- procedures for which usually incur some dependence on theoretical models -- leads immediately to determination of their luminosities using the formula $L = 4 \pi R^2 \Teff^{\ 4}$, making EBs excellent distance indicators. EB-based distances are available for open clusters as well as for nearby galaxies \citep[e.g.][]{Bonanos+06apj,North++10aa}. Another use of EBs is as tests and calibrators of stellar evolutionary theory. The predictions of theoretical models are required to match the measured values of mass, radius, \Teff\ and luminosity for two stars of the same age and chemical composition \citep[e.g.][]{Pols+97mn,Claret07aa}.

Many EBs contain stars which show intrinsic variability due to pulsations. Stochastic oscillations are observed in solar-type dwarfs, are present with much higher amplitudes in cool giant stars, and can be used to determine the density, radius and thus mass of the star to high precision \citep{Chaplin+14apjs}. Although of particular importance for solar-type dwarfs, stochastic oscillations have so far only been detected in the giant components of EBs. Other types of pulsations which have been found in conjunction with eclipses are $\delta$\,Scuti, $\gamma$\,Doradus, spB and tidally-induced oscillations. In this work I review recent results for these objects, then consider the future impact of current and forthcoming space missions.


\section{Red giants in eclipsing binaries}

The one type of system where properties from asteroseismology can currently be compared to those determined via the effects of binarity is that of eclipsing red giants. A single eclipse, 0.12\,mag deep and lasting 2.2\,d, was found in the Quarter 1 light curve of KIC\,8410637 \citep{Hekker+10apj} obtained by the \kepler\ satellite. Subsequent spectroscopic follow-up observations coupled with further eclipses found in the \kepler\ data have yielded accurate mass and radius measurements for the giant ($M_{\rm A} = 1.557 \pm 0.028$\Msun, $R_{\rm A} = 10.74 \pm 0.11$\Rsun) and its companion ($M_{\rm B} = 1.322 \pm 0.017$\Msun\ and $R_{\rm B} = 1.571 \pm 0.031$\Rsun), which pursue orbits of period 408.3\,d and eccentricity $e = 0.6864 \pm 0.0019$ \citep{Frandsen+13aa}. The properties of the giant from asteroseismology ($1.83 \pm 0.14$\Msun, $11.58 \pm 0.30$\Rsun; \citealt{Huber15conf}) are larger than the results from the EB analysis by $1.9\sigma$ and $2.7\sigma$, respectively, for reasons which are currently under investigation.

\citet{Gaulme+13apj} has found 12 more similar systems using \kepler\ data, with orbital periods ranging from 14.7\,d to 1058\,d. \citet{Gaulme+14apj} noticed that the giants in the shorter-period EBs do not show stochastic oscillations, to a detection limit far below the oscillation amplitudes expected for such objects. This suggests that tidal effects spin up the giant, causing magnetic fields which dissipate the acoustic modes. \citet{Gaulme+14apj} found that mode depletion occurred when the fractional radius of the giant was larger than 0.16--0.24.

HD\,181068 (KIC\,5952403) is a bright giant observed by \kepler. It is the primary component of a hierarchical triple system and is orbited by a short-period binary containing two K-dwarfs \citep{Derekas+11sci}. The two dwarfs are eclipsing on a 0.9\,d period, and themselves transit and are occulted by the G-type giant on a 45.5\,d period. The giant star shows no stochastic oscillations such as are found in single stars of similar mass and radius, but instead shows oscillations at multiples of the orbital frequency. \citet{Fuller+13mn} found four frequencies in the \kepler\ Quarter 1--11 data: all are related to the orbital frequency and at least two are tidal in origin. \citet{Borkovits+13mn} determined the full physical properties of this triple system by using eclipse timing variations in the short-period EB as a substitute for RVs of the two dwarfs, which are unavailable due to their relative faintness. Tidally-induced pulsations have previously been seen in HD\,174884 (CoRoT\,7758), an EB consisting of two unevolved B-stars \citep{Maceroni+09aa}.


\section{$\delta$\,Scuti stars in eclipsing binaries}

\begin{figure}[t]
\begin{center}
\includegraphics[width=0.8\textwidth]{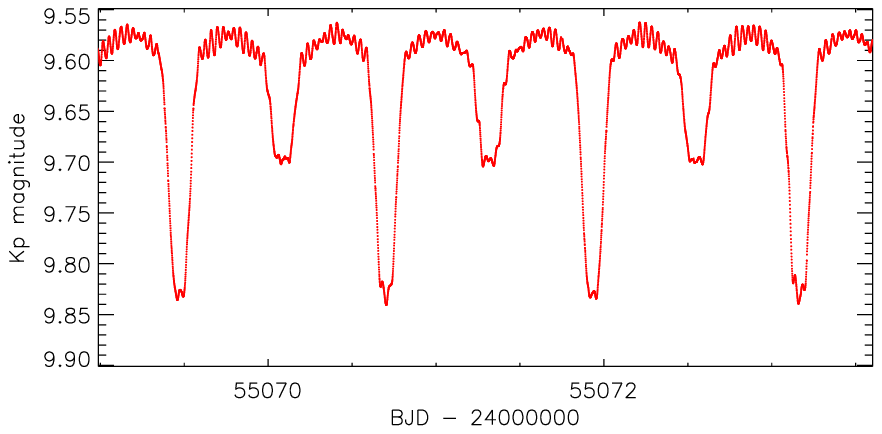}
\includegraphics[width=0.8\textwidth]{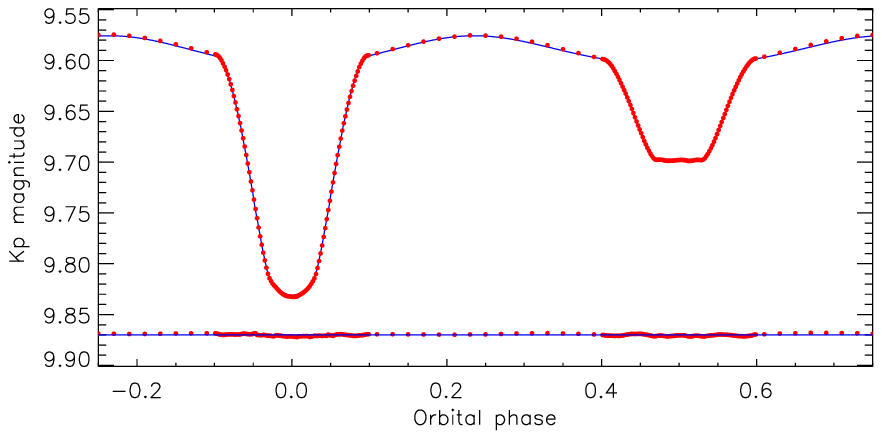}
\caption{\kepler\ light curve of the $\delta$\,Scuti EB KIC\,10661783 \citep{Me+11mn}. Top:
part of the short-cadence observations. Bottom: phased light curve after removal of the pulsations.}
\label{fig:1066}
\end{center}
\end{figure}

$\delta$\,Scuti stars are A- and F-type dwarfs which pulsate in non-radial $p$-modes with periods from 0.014\,d to 0.333\,d \citep{Grigahcene+10apj}. A substantial number of known $\delta$\,Scuti stars are members of EBs, including both detached and semi-detached systems. \citet{Mkrtichian+02aspc} defined the class of $\delta$\,Scuti stars in semi-detached systems, which have since been labelled as `oEA' (oscillating Algol) stars. Recent advances in the asteroseismic analysis of $\delta$\,Scuti stars have enabled the detection of large frequency separations analogous to solar-like oscillations but in a different pulsation regime \citep{Garcia+13aa}, representing a step towards utilising these pulsators as tests of stellar physics. \citet{Garcia+15xxx} have used EBs observed with \kepler\ and CoRoT to define a relation between the large frequency separation and mean density for $\delta$\,Scuti stars.

An early result from the \kepler\ satellite was the detection of eclipses and 68 pulsation frequencies in the light curve of KIC\,10661783 \citep{Me+11mn}. Follow-up spectroscopy by \citet{Lehmann+13aa} allowed determination of the masses and radii of the two stars. The secondary component appears to be a classic example of close-binary evolution, having a mass of $0.1913 \pm 0.0025$\Msun\ and a radius of $1.124 \pm 0.019$\Rsun. However, the spectroscopic mass ratio leads to a light curve model in which the two stars are currently detached and therefore not undergoing mass transfer. KIC\,10661783 may be an example of a semi-detached system which shows only episodic mass transfer.

Other examples include KIC\,3858884 \citep{Maceroni+14aa}, a detached EB with an orbital period of 26.0\,d and a high eccentricity ($e = 0.47$). Theoretical models require convective core overshooting to match the properties of this system. CoRoT 105906206 is another, for which \cite{Dasilva+14aa} required the inclusion of Doppler beaming to fit the light curve.


\section{$\gamma$\,Doradus stars in eclipsing binaries}

\begin{figure}[t]
\begin{center}
\includegraphics[width=0.8\textwidth]{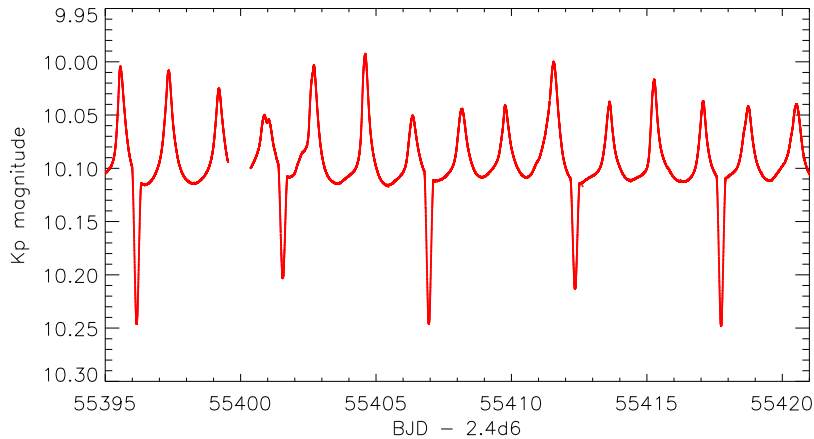}
\caption{\kepler\ light curve of KIC\,11285625, which shows eclipses and $\gamma$\,Dor pulsations.}
\label{fig:1128}
\end{center}
\end{figure}

\begin{figure}[t]
\begin{center}
\includegraphics[width=0.8\textwidth]{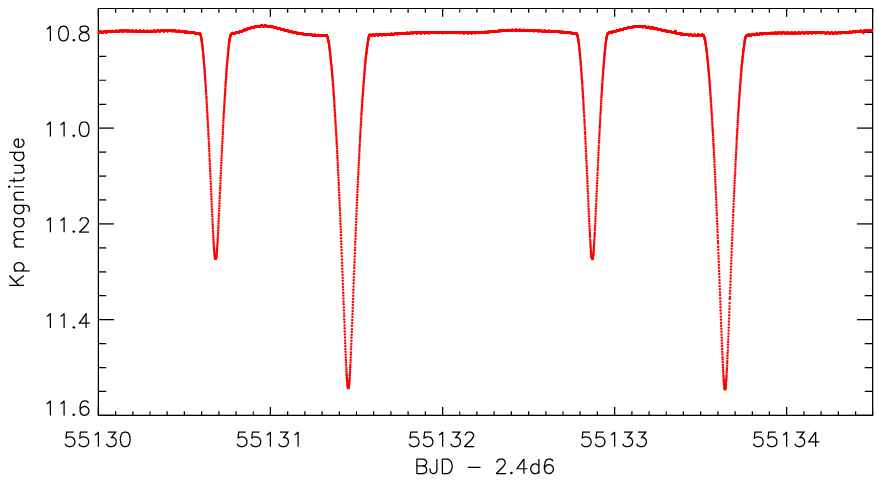}
\includegraphics[width=0.8\textwidth]{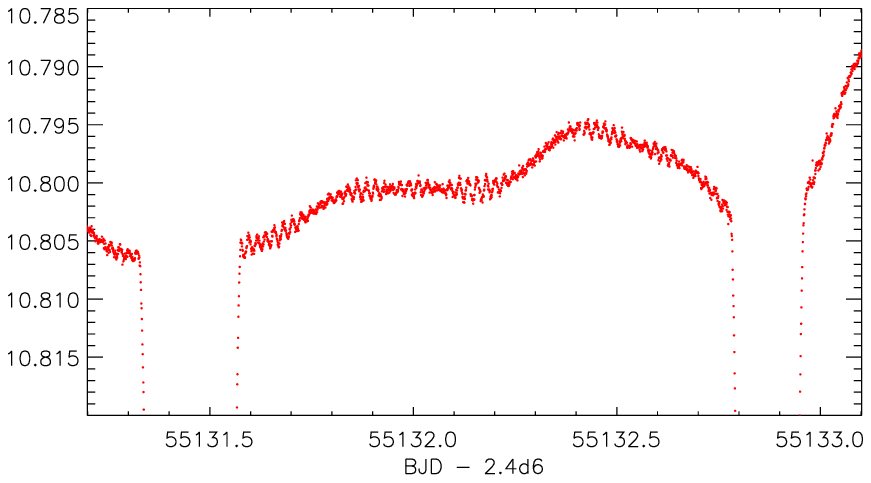}
\caption{\kepler\ light curve of KIC\,4544587, which shows eclipses,
$\delta$\,Scuti and $\gamma$\,Dor pulsations. Top:part of the short-cadence
observations. Bottom: close-up of an short stretch of data outside eclipse.}
\label{fig:0454}
\end{center}
\end{figure}

$\gamma$\,Dor stars are non-radial $g$-mode pulsators with typical periods of 0.4\,d to 3\,d \citep{Handler99mn} and spectral types of A7-F5 V-IV. The advent of high-precision photometry from space satellites has revealed that a large fraction are hybrids which show both $\gamma$\,Dor and $\delta$\,Scuti pulsations \citep{Grigahcene+10apj}. Recent advances in asteroseismology have enabled the detection of period spacings which hold the prospect of detailed seismic analyses of these objects \citep{Vanreeth+15aa,Kurtz+15mn}.

It was inevitable that some $\gamma$\,Dor stars would be found in EBs, and cases have indeed been identified from \kepler\ observations. \citet{Debosscher+13aa} presented a detailed analysis of KIC\,11285625, which shows 12\%-deep eclipses on a 10.8\,d period plus $g$-mode oscillations of similar amplitude (Fig.\,\ref{fig:1128}). Analysis of the \kepler\ data and ground-based radial velocity measurements allowed Debosscher et al.\ to determine the masses and radii of the two component of the EB to 1\%. Amplitude modulation and splitting at the rotational and orbital frequencies was also found from analysis of the pulsations.

Another example is KIC\,4544587 \citep{Hambleton+13mn}, an EB with an eccentric 2.2\,d period orbit and rapid apsidal motion. Whilst the eclipses dominate the \kepler\ light curve (Fig.\,\ref{fig:0454}), both $\delta$\,Scuti and $\gamma$\,Dor pulsations are visible on closer inspection. \citep{Hambleton+13mn} found 14 $g$-modes, most of which were harmonics of the orbital frequency, and 17 $p$-modes which were either affected by tides or occurring within the secondary star.


\section{Stochastic oscillations in eclipsing binaries}

\begin{figure}[t]
\begin{center}
\includegraphics[width=0.8\textwidth]{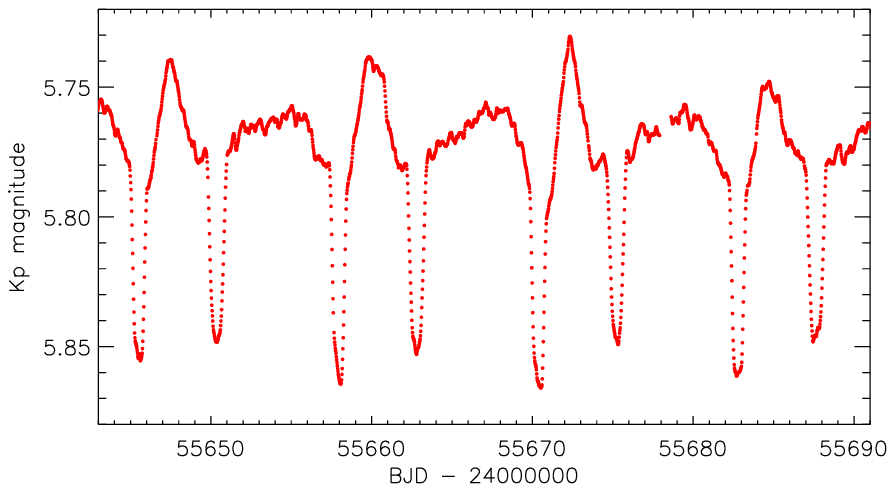}
\caption{Fragment of the \kepler\ short-cadence light curve of V380\,Cygni.}
\label{fig:v380}
\end{center}
\end{figure}

A major goal for space-based transit surveys such as \kepler\ and PLATO is the identification of potentially habitable planets. The widely accepted criteria for such planets is that they are Earth-like: rocky and orbiting within the habitable zone of their host star \citep[e.g.][]{Rauer+14exa}. The low masses and radii of such planets, coupled with their long orbital periods, makes them difficult to characterise. Their radii will be measured relative to those of their host stars, so it is important to obtain as much information as possible about these stars. Asteroseismology is an important component of such work \citep[e.g.][]{Christensen+10apj,Borucki+12apj}, and EBs offer an unrivalled opportunity to check the reliability of the asteroseismic results.

However, this has yet to be achieved because stochastic oscillations have not been detected in a solar-type eclipsing binary. The short-period EBs observed using \kepler\ suffer from tidal effects which cause rotational splitting and dissipation of the acoustic modes, making them undetectable in current data. The scarcity of longer-period EBs (orbital period greater than approximately 10\,d depending on the masses and evolutionary statuses of the components) means that no suitable candidates have been found for a comparison between asteroseismic properties and those measured from eclipses and radial velocities.

One notable success has been the detection of stochastic oscillations in the extremely bright ($V = 5.7$) totally-eclipsing binary V380\,Cygni \citep{Tkachenko+12mn,Tkachenko+14mn}, which is composed of a B1.5 giant and a B2 dwarf of known chemical composition \citep{Pavlovski+09mn}. Fig.\,\ref{fig:v380} shows part of the \kepler\ light curve of this system, in which can be seen primary and secondary eclipses, periastron brightening due to tidal effects, and stochastic variability. A Fourier transform of the \kepler\ light curve after subtraction of the effects of binarity shows some frequencies which are multiples of the orbital frequency and others which are not. The physical properties of the binary can only be matched by theoretical models with unrealistically high rotation and strong convective core overshooting, implying that much increased mixing is needed in these models.


\section{Present and future}

Although this review has showcased many results based on data from the \kepler\ satellite, the archive of \kepler\ data contains over 2700 EBs (Kirk et al., in press) of which the vast majority have not received significant attention. The CoRoT satellite also obtained over 163\,000 light curves including over 3000 EBs \citep{Moutou+13icar}, and the K2 mission is continuing to accumulate copious high-quality space-based data \citep{Howell+14pasp,Conroy+14pasp}. The near future holds the prospect of the TESS satellite, which will observe most of the sky for 27\,d intervals and thus greatly expand the database for bright EBs.

One of the remaining tasks is to study solar-type EBs for which stochastic oscillations can be detected. This necessarily requires space-based photometry in order to get observations with a high duty cycle, low scatter, and long time coverage. Each of the past and present space missions has at least one feature which compromises its ability to perform this experiment, such as low sampling rate (most of the \kepler\ targets), faintness of the stars observed (\kepler\ and CoRoT), or insufficient duration of observations (TESS, except for small patches of sky near the celestial poles).

The PLATO mission promises to solve all of these problems \citep{Rauer+14exa}. Scheduled for launch in 2024, it will observe hundreds of thousands of bright stars at high cadence and for time intervals up to several years. Extensive effort will be needed to cope with this inundation of data, and to procure follow-up observations from ground-based facilities. The work package structure for PLATO includes one dedicated to binary and multiple stars, for which the author is responsible, within the Complementary Science Work Package group. Although preparation of an extensive target list of known EBs will form part of this work package, it is likely that the majority of EBs observed by PLATO will be previously unidentified binaries selected for observation in one of the main transit search samples. These will, by design, be F, G, K and M dwarfs and the study of those showing eclipses promises to revolutionise our understanding of solar-type and low-mass stars.


\bibliographystyle{mn_new}

\end{document}